\documentclass[aps,prx,twocolumn,amsmath,amssymb]{revtex4}

\usepackage{amsmath, amsthm, amsfonts}
\usepackage{graphicx}
\usepackage{color,amssymb}
\usepackage{psfrag}
\usepackage{longtable}
\usepackage{dcolumn,epsfig}
\usepackage[normalem]{ulem}
\usepackage{color}
\usepackage{comment}

\newcommand{\abs}[1]{\left\vert#1\right\vert}
\newcommand{\be}{\begin{equation}}
\newcommand{\ee}{\end{equation}}
\newcommand{\ba}{\begin{eqnarray}}
\newcommand{\ea}{\end{eqnarray}}

\newcommand{\ban}{\begin{eqnarray*}}
\newcommand{\ean}{\end{eqnarray*}}

\begin{document}

\title{Observation of the Three-Mode Parametric Instability}
\author{X. Chen}
\affiliation{School of Physics, University of Western Australia, WA 6009, Australia}
\author{C.Zhao}
\affiliation{School of Physics, University of Western Australia, WA 6009, Australia}
\author{S. Danilishin}
\affiliation{School of Physics, University of Western Australia, WA 6009, Australia}
\author{L. Ju}
\affiliation{School of Physics, University of Western Australia, WA 6009, Australia}
\author{D. Blair}
\affiliation{School of Physics, University of Western Australia, WA 6009, Australia}
\author{H. Wang}
\affiliation{Department of Physics, Beijing Normal University, Beijing 100875, China}
\author{S. P. Vyatchanin}
\affiliation{Faculty of Physics, M.V. Lomonosov Moscow State University, Moscow 119992, Russia}
\author{C. Molinelli, A. Kuhn, S. Gras$^{*}$, T. Briant, P.-F. Cohadon, and A. Heidmann}

\affiliation{Laboratoire Kastler Brossel, UPMC - Sorbonne Universit\'es, CNRS, ENS - PSL Research University, Coll\`ege de France, 75005 Paris, France\\$^{*}$ Now at LIGO Laboratory, MIT, Boston, USA
}

\author{I. Roch-Jeune}
\affiliation{Institut d'Electronique, de Micro\'electronique et de
Nanotechnologie, UMR 8520 CNRS, 59652 Villeneuve d'Ascq, France}

\author{R. Flaminio, C. Michel, L. Pinard}
\affiliation{Laboratoire des Mat\'{e}riaux Avanc\'{e}s, IN2P3/CNRS, Universit\'{e} de Lyon, F-69100 Villeurbanne, France}

\date{\today}

\begin{abstract}

Three-mode parametric interactions occur in triply-resonant optomechanical systems: photons from an optical pump
mode are coherently scattered to a high-order mode by mechanical motion of the cavity mirrors, and these modes resonantly interact via radiation pressure force when certain conditions are met. Such effects are predicted to occur in long baseline advanced gravitational-wave detectors.  They can pump energy into acoustic modes, leading to parametric instability, but they can also extract acoustic energy, leading to optomechanical cooling. We develop a large amplitude model of three-mode interactions that explains the ring-up amplitude saturation after instability occurs. We also demonstrate both radiation-pressure cooling and mechanical amplification in two different three-mode optomechanical systems, including the first observation of the three-mode parametric instability in a free-space Fabry-Perot cavity. The experimental data agrees well with the theoretical model. Contrary to expectations, parametric instability does not lead to loss of cavity lock, a fact which may make it easier to implement control techniques to overcome instability.
\end{abstract}

\maketitle

In 2001, Braginsky \emph{et al.}  predicted three-mode parametric instability in advanced gravitational-wave detectors \cite{Braginsky2001, Braginsky2002}, which would be caused by coincident frequency matching and mode shape matching between mirror mechanical modes and high-order optical modes. The high mechanical and optical mode density of these systems makes it likely that such interactions occur accidentally. Subsequently detailed modeling \cite{zhao2005,KellsMod, Evans} verified the predictions and experimental tests on suspended optical cavities \cite{zhaoPRA2008,zhaoPRA2011} demonstrated three-mode interactions below the instability threshold.

Three-mode interactions mimic a two-level atomic system \cite{Grudinin}, in which parametric instability is analogous to the creation of a phonon laser. Bahl \emph{et al.}  emphasize that this phenomenon is a macroscopic realization of Brillouin scattering \cite{Bahl,Brillouin1965}. The parametric gain can in principle be tuned to cause either mode amplification (which can lead to instability) or mode cooling.

In three-mode interaction, both the injected optical mode and the scattered mode are resonant in the cavity.  This enhances the optomechanical coupling, so that the input power required for parametric instability is reduced compared to two-mode parametric interactions which require detuning from resonance and higher input power \cite{Kippenberg2010}.

 Well understood two-mode parametric instabilities have been observed in a suspended cavity \cite{MIT1}, and in various solid state optical microresonators \cite{Kippenberg1, Kippenberg2, Kippenberg3, Carmon, CarmonBahl, Carmom2007, Ma}. Three-mode instability has been observed in relatively low quality factor solid-state resonators \cite{Grudinin2009, Tomes2009, SAW2009, SAW2011, Kippenberg4} and in a microwave system \cite{Tobar1}, but has not been reported in free-space optical cavities. The challenge can be met either by using very high optical power in large scale optical cavities such as Advanced LIGO \cite{aLIGO} or Advanced Virgo \cite{aVirgo} detectors now under construction, or at low power in table-top cavities with suitable mode structure and a low mass high quality factor mechanical resonator.

  In this paper we present two free-space cavity configurations suitable for investigating three-mode interactions and parametric instability. One is a cavity coupled to a silicon bridge microresonator, which has already been used for two-mode parametric cooling and instability experiments \cite{NatureLKB}. The second is a free-space cavity with an intracavity membrane \cite{Zwickl2008}. Both systems are designed to have suitable optical mode structures which can be matched to the mechanical mode structures. The membrane system has the lowest mass, and has enabled the first observation of three-mode parametric instabilities  in a system somehow similar to gravitational-wave detectors.

We first present a summary of the conventional theory of three-mode parametric interactions. We proceed by describing a large-amplitude model that would be valid after parametric instability occurs. Numerical simulation of the temporal evolution of the intracavity fields and the mechanical displacement are given. We then describe the experiment with a bridge resonator, which allows us to demonstrate the resonant character of both amplification and cooling processes. Finally, we describe a membrane experiment, which has demonstrated for the first time parametric instability in a free-space cavity. The latter experimental results reveal that the instability does not lead to the loss of cavity locking and are in excellent agreement with the large-amplitude model. The implications of our results for advanced gravitational-wave detectors are briefly discussed.

\section{Simple theory of Three-Mode Parametric Interactions}

\begin{figure}[t!]
\centering
\includegraphics[width=0.5\textwidth]{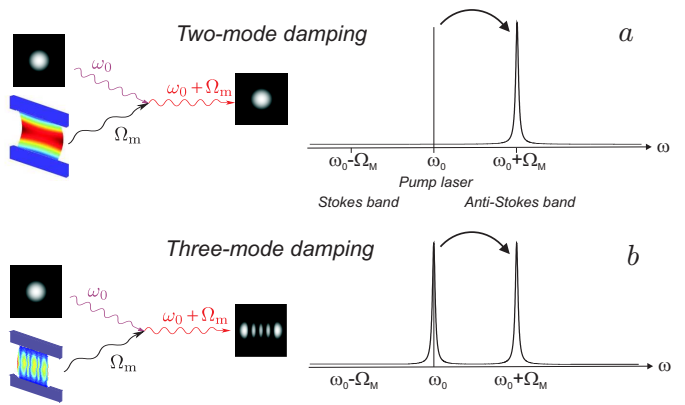}
\caption{Radiation-pressure damping seen as a scattering process.
 \emph{a}: \emph{Two-mode damping}. The incident photon at $\omega_0$ is scattered to the same optical mode, with a
frequency offset $\Omega_{\rm m}$ w.r.t. the incident photon. The
cavity is detuned in order to favour the anti-Stokes process, and
the pump laser is  far detuned from the resonance.  \emph{b}: \emph{Three-mode
damping}. The photon is scattered to a different optical mode of the
cavity. Both the pump laser and the anti-Stokes band can be
simultaneously resonant with the cavity. The vibration profile of
the mechanical mode must match the shape of output optical mode to achieve strong coupling.}
\label{figure:Principe3Modes}

\end{figure}

We will first review the theory of two- and three-mode interactions in the context of cooling, amplification and instability. Fig. \ref{figure:Principe3Modes} presents a cartoon of the two- and three-mode interactions for the case of the anti-Stokes damping process.
An incident laser (angular frequency
$\omega_0=2\pi/\lambda$) is scattered by a moving mirror (resonance frequency $\Omega_{\rm m}/2\pi$), creating two sidebands: Stokes (at
$\omega_0-\Omega_{\rm m}$) and anti-Stokes ($\omega_0+\Omega_{\rm m}$).
In a cavity cooling experiment, the anti-Stokes band has to be favoured by
tuning the cavity close to $\omega_0+\Omega_{\rm m}$ to damp and effectively cool the resonator motion.
For the two-mode interaction case  shown in Fig. \ref{figure:Principe3Modes}a, the final number of
mechanical quanta is related to the residual Stokes processes, which can
be negligible as long as one operates in the resolved sideband
regime \cite{Kippenberg,TheorySelfCool,TheorySelfCool2}
where the cavity linewidth is small compared to $\Omega_{\rm m}$. This is the
regime in which the quantum ground state of mechanical resonators has been recently demonstrated \cite{QGS_Teufel,QGS_Painter,QGS_TJK}. For such two-mode resolved sideband cooling, the incident laser beam must be detuned very far from the optical resonance, and high incident power
$P_{\rm in}$ must be used to achieve efficient cooling. Most of
the light is reflected by the cavity, and is not coupled to resonator motion.

Three-mode systems overcome the problem of large offsets and poor optical coupling by providing a second optical mode tuned to the sideband frequencies. In this case, the mechanical
mode of the moving mirror is coupled to not just one but two different optical cavity modes. The incident laser beam (pump
mode $\omega_{\rm p}$) and the anti-Stokes sideband (target higher-order mode $\omega_{\rm s}$) can then be
simultaneously resonant with the cavity, while still keeping the
Stokes band far detuned. In this situation, the cooling process
appears as a scattering process of laser photons to a
higher-frequency mode by mechanical phonon absorption.
The maximum efficiency is obtained when the laser is resonant with the pump mode and when
the   frequency offset $\Delta\omega=\omega_{\rm s}-\omega_{\rm p}$ between the target mode and the pump
is close to $\Omega_{\rm m}$, as shown in Fig. \ref{figure:Principe3Modes}b.

We have so far emphasized the Anti-Stokes case because it is the most explored experimentally
but the Stokes sideband can obviously be tuned to resonance as well
with a similar configuration. In this case, one expects to observe three-mode
amplification with  the energy flow from the pump field to the
mechanical resonator. For sufficient gain, this can lead to a three-mode instability.

The effective damping rate $\Gamma_{\rm eff}$ of the mechanical resonator
(mass $m$, mechanical quality factor $Q$ and intrinsic damping $\Gamma_{\rm m}=\Omega_{\rm m}/Q$) by the Stokes process
is related to the
parametric gain $\mathcal{R}$ which fully characterizes the three-mode parametric interactions:
\begin{equation}
\frac{\Gamma_{\rm eff}}{\Gamma_{\rm m}}=1-{\cal R}.
\label{eq:gamma-eff}
\end{equation}
The parametric gain is given by
\begin{equation}
{\cal R} = {\cal R}_+-{\cal R}_-,
\end{equation}
where
\begin{equation}
{\cal R}_+=\frac{16}{\pi \lambda c }
\frac{\mathcal{F}_{\rm p}\mathcal{F}_{\rm t}Q P_{\rm in}\,\Lambda}{m\Omega_{\rm m}^2} \,\left(\frac{1}{1+\left(\omega_{\rm p}-\Omega_{\rm m}-\omega_{\rm s}\right)^2/\Omega_{\rm cav}^2}\right)
\label{eq:gainStokes}
\end{equation}
is the gain for the Stokes process and
\begin{equation}
{\cal R}_-=\frac{16}{\pi \lambda c }
\frac{\mathcal{F}_{\rm p}\mathcal{F}_{\rm t}Q P_{\rm in}\,\Lambda}{m\Omega_{\rm m}^2} \,\left(\frac{1}{1+\left(\omega_{\rm p}+\Omega_{\rm m}-\omega_{\rm s}\right)^2/\Omega_{\rm cav}^2}\right),
\label{eq:gainantiStokes}
\end{equation}
is the gain for the anti-Stokes process.
$\mathcal{F}_{\rm p}$ and  $\mathcal{F}_{\rm t}$ are the optical finesses for the pump and target
modes of the cavity, $\Omega_{\rm cav}/2\pi$ the optical
bandwidth of the scattered mode, and $\Lambda$ the spatial overlap between mechanical mode
$u$ and optical modes $v_{\rm p}$ and $v_{\rm s}$ (see Sec. \ref{secModel}). $\mathcal{R}$  is negative for cooling, positive for amplification and larger than 1 for instability \cite{Braginsky2001}.

\section{Large-amplitude model\label{secModel}}
All prior analysis of three-mode interactions have assumed  small amplitude \cite{Braginsky2001,Kells,Braginsky2002,Juli3mode,Strigin, Evans}. While appropriate for obtaining instability criteria, this approach is no longer relevant once the instability threshold is passed and the loss of fundamental mode power through scattering into high-order mode becomes large.
We develop a large-amplitude theoretical model, which is valid for parametric gain $\mathcal{R}>1$.  It shows that the interaction will reach a saturated steady state.
We also present the results of a numerical simulation, which will be used to interpret our experimental results in Sec. \ref{secExpUWA}. We assume here that the Stokes mechanism of the three-mode interaction is resonant (i.e. $\Delta\omega=+\Omega_{\rm m}$).

\subsection{Interaction Hamiltonian}

The Hamiltonian for the whole system is:
\ba
& &\hat{{H}} = \hat{H}_{\rm p}+\hat{H}_{\rm s}+\hat{H}_{\rm m}+\hat{H}_{\rm int}+\hat{H}_{\rm drive}\,,\\
& &\hat{{H}}_{i} = \hbar\omega_{i}\hat{a}^{\dag}_{i}\hat{a}_{i}\,,\quad (i=\rm p,\rm s)\,\\
& &\hat{{H}}_{\rm m} = \hbar\Omega_{\rm m} \left(\hat{b}^{\dag}_{\rm m}\hat{b}_{\rm m}+\frac{1}{2}\right)\,,\\
& &\hat{H}_{\rm drive} = i\,\hbar\sqrt{2\gamma_0}\, A_{\rm in}\left(\hat{a}^{\dag} {\rm e}^{-i \omega_0 t}-\hat a {\rm e}^{i \omega_0 t} \right)\,,
\ea
where $A_{\rm in} = \sqrt{P_{\rm in}/\hbar \omega_0}$, $P_{\rm in}$ is the input power, $\gamma_0$ is the pump field amplitude damping rate.
We assume in the following that the pump has the same frequency as the ${\rm TEM}_{00}$ mode$: \omega_0 = \omega_{\rm p}$.

If we define the effective mechanical vibration by its amplitude $\hat x_{\rm m}=(\hbar/2m\Omega_{\rm m})^{1/2}(\hat b^{\dagger}_{\rm m}+\hat b_{\rm m})$,  the interaction between the mechanical mode, the pump mode and the scattered mode (designated by subscripts ${\rm p}$ and ${\rm s}$ respectively) is described by the Hamiltonian  $\hat{H}_{\rm int}$ \cite{Vitali2011}:
\be\label{eq:hint}
\hat{H}_{\rm int}=-\hbar(\hat b^{\dagger}_{\rm m}+\hat b_{\rm m}) \sum_{i,j} G_{ij}\hat a^{\dagger}_i \hat a_j\,,\quad(i,j = \rm p,\rm s)\,,
\ee
with a coupling strength $G_{ij}$:
\be
\label{eq:coupling}
   G_{ij} = \frac{2}{L} \sqrt{\frac{\hbar \omega_i \omega_j}{2 m \Omega_{\rm m}}\Lambda_{ij}}\,,\quad (i,j=\rm p,\rm s)\, \\
\ee
where $L$ is the cavity length and $m$ is the effective mass of the mechanical mode. The overlap factor between the optical modes $\hat{a}_{\rm p}$, $\hat{a}_{\rm s}$ and the mechanical mode $\hat{b}_{\rm m}$ is defined as:
\be
    \Lambda_{ij}=\left(\int \mathrm{d}^2\mathbf{r}_\bot \; u(\mathbf{r}_\bot)\,v_i(\mathbf{r}_\bot)\, v_j(\mathbf{r}_\bot)\right)^2,
\label{eq:overlap}
\ee
where $(i,j=\rm p,\rm s)$. $u$ is the normalized mechanical mode profile, $v_{\rm p}$ and $v_{\rm s}$ the normalized optical mode profiles for the pump and scattered modes,
$\mathbf{r}_\bot$ being the transverse coordinate on the resonator surface \cite{Braginsky2001}.

Three-mode amplification is only significant when three-mode spatial overlap $\Lambda_{ij}$ is large. No interaction can take place without a proper mode shape matching, and likewise the optical signal detection must have the appropriate mode shape sensitivity. Assuming the cavity
is pumped with a Gaussian profile mode TEM$_{00}$, the vibration profile $u$ must have a non-zero
match to the output mode structure, corresponding to a finite value for $\Lambda_{ij}$.

\subsection{Equations of motion}

We separate the slow amplitude variation from the fast oscillation term by making the following transformation: $\hat{b}_{\rm m} \to b_{\rm m} {\rm e}^{-i\Omega_{\rm m} t}$, $\hat{a}_{\rm p} \to a_{\rm p} {\rm e}^{-i\omega_{\rm p} t}$ and $\hat{a}_{\rm s} \to a_{\rm s} {\rm e}^{-i\omega_{\rm s} t}$. Taking into account the damping rates $\Gamma_{\rm m}$, $\gamma_{\rm p}$ and $\gamma_{\rm s}$ of each degree of freedom, we obtain the following equations of motion:
\ba
\label{eq:PIfull_eqs}
&&\dot a_{\rm p} + \gamma_{\rm p} a_{\rm p} - i G_{\rm ps} a_{\rm s}(b_{\rm m}+b_{\rm m}^*{\rm e}^{2i\Omega_{\rm m} t}) = \sqrt{2 \gamma_{\rm p}}A_{\rm in} \,\,\\
&&\dot a_{\rm s} + \gamma_{\rm s}a_{\rm s} - i G_{\rm ps}a_{\rm p}(b_{\rm m} {\rm e}^{-2i\Omega_{\rm m} t}+b_{\rm m}^*)=0 \,\,\\
&&\dot b_{\rm m} + \Gamma_{\rm m} b_{\rm m} - i G_{\rm ps}\left(a_{\rm p}^* a_{\rm s} {\rm e}^{2i\Omega_{\rm m} t}+a_{\rm s}^* a_{\rm p}\right)=0\,.
\ea
The equations of motion cannot be solved exactly analytically. The simulation of the whole parametric instability process with Finite-difference time-domain (FDTD) method is shown in section \ref{sec:simulation}. We are interested in the mechanical amplitude at the resonance frequency $\Omega_{\rm m}$, so in sections \ref{sec:ringup} and \ref{sec:steadystate}, we only consider the slowly evolving part of the amplitude and the high-frequency oscillating terms are omitted.

\subsection{Numerical Simulation of Parametric instability}\label{sec:simulation}
To have a better understanding of the whole parametric instability process, we have performed a FDTD simulation. To reduce artifacts, we solve the equations of motion with the Runge-Kutta method \cite{Press1992}.
To reduce computation time, we rewrite the equations of motion to have three variables in the following form:
\ba
\label{eq:PIfdtd_eqs2}
\dot a_{\rm p}&=&-\gamma_{\rm p} a_{\rm p} + i G_{{\rm ps}}a_{\rm s}{\rm e}^{i\Omega_{\rm m} t}\times 2{\rm Re}(b_{\rm m} {\rm e}^{-i\Omega_{\rm m} t}) + \sqrt{2 \gamma_{\rm p}}A_{\rm in}\,,\nonumber \\
\dot a_{\rm s}&=&-\gamma_{\rm s}a_{\rm s} + iG_{{\rm ps}}a_{\rm p} {\rm e}^{-i\Omega_{\rm m} t}\times 2{\rm Re}(b_{\rm m} {\rm e}^{-i\Omega_{\rm m} t})\,,\nonumber \\
\dot b_{\rm m}&=&-\Gamma_{\rm m} b_{\rm m} + iG_{{\rm ps}}{\rm e}^{i\Omega_{\rm m} t}\times 2\textrm{Re} \left(a_{\rm p}^* a_{\rm s} {\rm e}^{i\Omega _{\rm m} t}\right)\,.
\ea
 The numerical values used for the simulation correspond to the experiment described in Sec. \ref{secExpUWA}. The initial state is set as:
\be
a_{\rm p}(0)=0,\, a_{\rm S}(0)=0, b_{\rm m}(0)= \sqrt{k_{\rm B} T/(\hbar \omega_{\rm m})},
\ee
so the system starts with the mechanical resonator thermal noise level and no light resonant in the cavity. In reality, the initial conditions are determined by the thermal distribution of the mode amplitude. Fluctuations in this amplitude (assumed here to have its mean value) lead to small changes on the effective time axis. Because the amplitudes of concern in this paper are large compared with the thermal amplitude, we also neglect the thermal driving of $b_{\rm m}$ in Eq. (\ref{eq:PIfdtd_eqs2}).

\begin{figure}[t]
\centering
\includegraphics[width=0.5\textwidth]{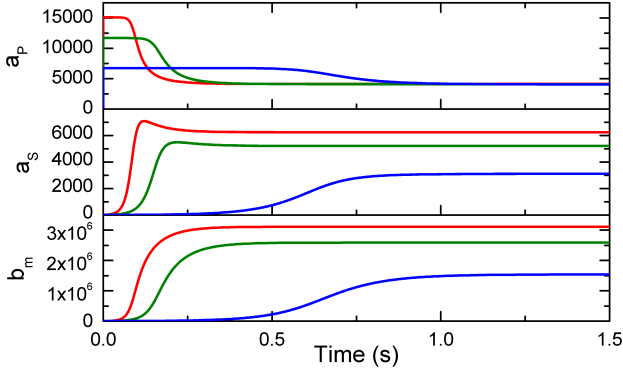}
\caption {Simulation results for the TEM$_{00}$ mode amplitude ($a_{\rm p}$), the TEM$_{20}$ scattered mode amplitude ($a_{\rm s}$) and the mechanical motion amplitude ($b_{\rm m}$). The input powers used here are $5\, \mu{\rm W}$ (black curves), $15\, \mu{\rm W}$ (blue curves) and $25\, \mu{\rm W}$ (red curves).The system acts to control the pump power in the cavity to a value which is independent of the incident power, by scattering into the mechanical and transverse modes.}
\label{fig:simulation2}
\end{figure}

The simulation results obtained with an input TEM$_{00}$ power of $5\, \mu{\rm W}$, $15\, \mu{\rm W}$ and $25\, \mu{\rm W}$  are shown on Fig. \ref{fig:simulation2}.
When $a_{\rm p}$ reaches a threshold value, the parametric amplification process starts and both $a_{\rm s}$ and $b_{\rm m}$ start growing. This process goes on until saturation is reached. In the saturation regime, the intracavity pump amplitude $a_{\rm p}$ is precisely independent of the pump power (see Fig. \ref{fig:simulation2}, top curve). The excess optical energy is channeled to the scattered optical amplitude $a_{\rm s}$ and the mechanical motion amplitude $b_{\rm m}$. For our parameters, above threshold, the typical scattered circulating power is a few mW and the typical displacement amplitude of the order of $10^{-10}$ m (see Fig. \ref{figure:steady}).

\subsection{Analytical study of the initial ring-up}\label{sec:ringup}
Here we only consider the slow variation terms and the equations of motion can be reduced to the following form:
\ba
\label{eq:PIfdtd_eqs3_a0}
\dot a_{\rm p} &=& -\gamma_{\rm p} a_{\rm p} + i G_{\rm ps}a_{\rm s}b_{\rm m} + \sqrt{2 \gamma_{\rm p}}A_{\rm in}\,,\\
\label{eq:PIfdtd_eqs3_as}
\dot a_{\rm s} &=& -\gamma_{\rm s}a_{\rm s} + i G_{ \rm ps}a_{\rm p} b_{\rm m}^*\,,\\
\label{eq:PIfdtd_eqs3_bm}
\dot b_{\rm m} &=& -\Gamma_{\rm m} b_{\rm m} + i G_{\rm ps}a_{\rm s}^* a_{\rm p}\,.
\ea
Combining the above equations leads to:
\be
\label{eq:4.24}
\ddot{b}_{\rm m} = -(\Gamma_{\rm m} +\gamma_{\rm s})\dot{b}_{\rm m} +G_{\rm ps}^2 a_{\rm p}^2b_{\rm m} -\gamma_{\rm s} \Gamma_{\rm m}\,.
\ee
This is an ordinary differential equation with solution in the form:
\be
\label{eq:4.25}
b_{\rm m} = B_{\rm m} {\rm e}^{\Gamma t}\,,
\ee
where $B_{\rm m}$ is the initial amplitude at $t=0$ and $\Gamma$ is the ring-up rate. By substituting Eq. \eqref{eq:4.25} into Eq. \eqref{eq:4.24} and solving the quadratic equation, we get:
\be
\Gamma = \frac{1}{2}\left(\sqrt{\left(\gamma_{\rm s}-\Gamma_{\rm m}\right)^2+ 8 G_{\rm ps}^2 P_{\rm in}/\left(\gamma_{{\rm p}} \hbar\,\omega_{\rm p}\right)}- \gamma_{\rm s}-\Gamma_{\rm m}\right)
\ee

Assuming $\gamma_s\gg \Gamma_{\rm m}$, which is true for most experimental situations, we then have:
\be
\Gamma \simeq \frac{1}{2}\left(\sqrt{\gamma_{\rm s}^2+ 8 G_{\rm ps}^2 P_{\rm in}/\left(\gamma_{{\rm p}} \hbar\,\omega_{\rm p}\right)}- \gamma_{\rm s}\right).\label{EqGamma}
\ee

From the above equations we can obtain the parametric gain:
\be
\mathcal{R}=\frac{2 G_{\rm ps}^2}{ \gamma_{\rm p} \gamma_{\rm s}\Gamma_{\rm m}}\frac{P_{\rm in}}{\hbar \omega_{\rm p}}.
\ee
The threshold input power for parametric instability ($\mathcal{R}\geq1$) is thus:
\be
\label{eq:threshold}
P_{\rm th} = \frac{\hbar \omega_{\rm p} \gamma_{\rm p} \gamma_{\rm s}\Gamma_{\rm m}}{ 2G_{\rm ps}^2}
\ee
and the ring-up time constant $\tau = 1/\Gamma$.

\subsection{Properties of the steady state}\label{sec:steadystate}
When the system reaches equilibrium, $\dot a_{\rm p} =0$,  $\dot a_{\rm s} =0$ and  $\dot b_{\rm m} =0$.
The equations of motion are then:
\ba
  & &\gamma_{\rm p} a_{\rm p} - i G_{{\rm ps}}a_{\rm s}b_{\rm m} = \sqrt{2 \gamma_{\rm p}}A_{\rm in}\,,\label{eq:steady_a0}\\
  & &\gamma_{\rm s}a_{\rm s}^* + i G_{{\rm ps}}a_{\rm p}^* b_{\rm m} = 0\,,\label{eq:steady_as}\\
  & &\Gamma_{\rm m} b_{\rm m} - iG_{{\rm ps}}a_{\rm s}^* a_{\rm p} = 0\,.\label{eq:steady_bm}
\ea

\begin{figure}[t]
\begin{center}\includegraphics[width=0.5\textwidth]{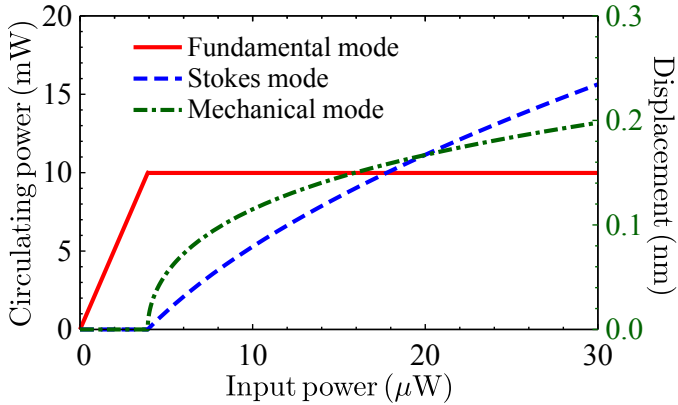}
\caption{Steady-state behaviour of three-mode parametric instability.  As soon as the cavity pump circulating power exceeds the instability threshold, the power remains constant for all input powers, while the transverse mode and the mechanical mode amplitude increase monotonically.
 \label{figure:steady}}
\end{center}\end{figure}

Substituting Eq. \eqref{eq:steady_as} into Eq. \eqref{eq:steady_bm}, we get:
\be \label{eq:steady_sol_a0}
 \frac{G^2_{{\rm ps}} \abs{a_{\rm p}}^2}{\Gamma_{\rm m} \gamma_{\rm s}}=1\,.
\ee
This means that when the system has reached its equilibrium state, the ${\rm TEM}_{00}$ intracavity power is independent of the input power,
which is a typical behaviour in many laser systems.
Substituting Eq. \eqref{eq:steady_sol_a0} into Eq. \eqref{eq:steady_as}, we get the corresponding scattered power:
\be
\label{eq:steady_sol_as}
\abs{a_{\rm s}}^2=\frac{\Gamma_{\rm m}}{\gamma_{\rm s}}\abs{b_{\rm m}}^2,
\ee
and using Eqs. \eqref{eq:steady_a0}, \eqref{eq:steady_as} and \eqref{eq:steady_sol_a0}, we finally obtain the steady-state mechanical motion amplitude:
\be
\abs{b_{\rm m}}^2=-\frac{\gamma_{\rm p} \gamma_{\rm s}}{G_{{\rm ps}}^2}+\sqrt{\frac{2  \gamma_{\rm p} \gamma_{\rm s}}{\Gamma_{\rm m} G_{{\rm ps}}^2}\frac{P_{\rm in}}{\hbar \omega_{\rm p}}}.
\ee
Fig. \ref{figure:steady} shows the steady-state amplitude of the cavity mode, the higher-order mode and the mechanical mode as a function of input power.

We have conducted two table-top free-space cavity experiments for three-mode parametric interaction study. To observe the three-mode interaction, two relevant optical modes have to be simultaneously resonant inside the cavity, while the frequency difference between the two optical modes must be tuned to the mechanical mode frequency to obtain maximum parametric gain \cite{Braginsky2001}.  The resonance frequencies of the mechanical modes of a typical table-top free-space optomechanical resonator  span from a few hundreds of kHz to a few MHz, while the free spectral ranges of cm-scale  optical cavities usually lie in the GHz-range.  The two optical modes therefore need to be almost degenerate (compared with the scale of the free spectral range) and to be tuned with a relative precision of $\sim 10^{-6}$ . This can be achieved with a careful choice of the mirror radii of curvature and of the cavity length.

The second great challenge in these experiments is to ensure a significant spatial overlap between the mechanical mode and the optical modes. The two table-top experiments described below achieve appropriate conditions for investigating three-mode parametric cooling, amplification and parametric instability.  One experiment uses  a silicon bridge resonator in a semi-confocal cavity, the other uses a membrane-in-the-middle configuration.

\section{Semi-confocal cavity with silicon bridge resonator: parametric cooling and amplification}

Our first experiment exactly corresponds to the actual setup first described by Braginksy \emph{et al.} \cite{Braginsky2001}: a single-ended linear Fabry-Perot cavity with a moving end mirror,
whose motion induces sidebands that can be amplified by the cavity.

\subsection{Experimental setup: design and characterization}

\begin{figure}[b!]
\begin{center}\includegraphics[width=7.5cm]{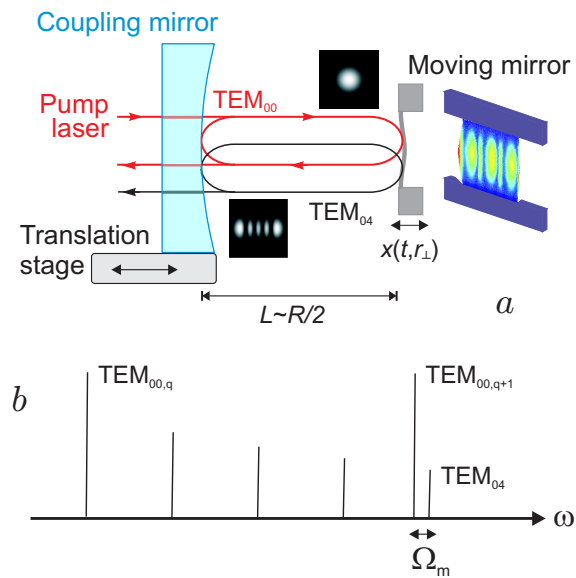}
\caption{Experimental implementation of three-mode coupling with a silicon resonator and a semi-confocal cavity. \emph{a}: The
 three-mode optomechanical system. A TEM$_{00}$ pump field is sent into the resonant
 cavity. This field is resonantly scattered to the TEM$_{04}$ mode by the
 moving mirror. The cavity length is tuned in order to set one of the motion-induced sidebands at
 resonance.
 \emph{b}: Mode spectrum of the cavity close to the semi-confocal working point.
 \label{figure:Cavite3Modes}}
\end{center}\end{figure}

The schematic experimental cavity and mode structure of the cavity is shown in Fig. \ref{figure:Cavite3Modes}. The moving mirror is a  $1\,\mathrm{mm}\times
800\,\mu\mathrm{m}\times 30\,\mu\mathrm{m}$ silicon doubly-clamped beam
\cite{NatureLKB,PRL_Arcizet}. Such a micromirror has a number of
mechanical modes with appropriate vibration profiles (see Fig.  \ref{fig:results}) and mechanical resonance frequencies
close to 5 MHz. A simple characterization setup using a network analyzer, local electrostatic actuation and a Michelson interferometer to probe the mirror motion
allows us to map the vibration profiles and therefore identify the different vibration modes.   The corresponding overlap factors $\Lambda_{\rm ps}$ can then be computed, with Gaussian modes TEM$_{00}$ and TEM$_{04}$ as pump and scattered modes, respectively.

The frequency resonance condition is reached with a cavity close to
the semi-confocal configuration, with the moving
mirror used as a nearly flat end mirror. The input mirror is concave,
with a 50-mm radius of curvature. As the coating thickness ($\simeq 5\,\mu$m) is not negligible compared to the resonator thickness, the coating process
yields tension within and bends the resonator, slightly changing the optical frequency resonance condition. The cavity has a
finesse of 30,000  for the TEM$_{00}$ pump mode, 24,000  for the
TEM$_{04}$ scattered mode, with a corresponding cavity bandwidth $\Omega_{\rm cav}/2\pi$ around 80 kHz. The
resonance frequency offset $\Delta\omega/2\pi$ scales linearly with the cavity length
$L$, with a measured rate of 250 kHz/$\mu$m close to the
degeneracy point between the TEM$_{00}$ and TEM$_{04}$ modes.

Demonstration of three-mode interactions requires the tuning of the frequency difference
$\Delta\omega$ with a precision that is small compared with the cavity bandwidth
$\Omega_{\rm cav}$.  This requires the cavity length to be controlled at the 100-nm
level. To change the cavity length, we have used a stepper motor to set either the
Stokes or the anti-Stokes band close to resonance, and a piezoelectric transducer (PZT) for finer
displacement tuning in the vicinity of the resonances. When operating an optical cavity close to a degeneracy point, high mechanical stability is also required, especially for this high-finesse moving mirror cavity.

For each set of experiment, we first measure an optical spectrum of the cavity, similar to the one presented in Fig. \ref{fig:gap} (for the other experimental setup). This, together with an identification of the transverse mode corresponding to each resonance peak with a CCD camera, allows to measure the exact optical detuning for a given length of the cavity, which has proven stable over the typical duration of a complete experiment. We then measure the noise spectra for different cavity detunings.

\subsection{Experimental results with the semi-confocal cavity}

\begin{figure}[b]
\includegraphics[width=7.5cm]{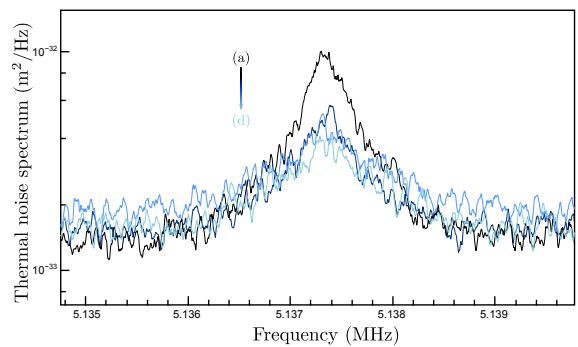}
\caption{Three-mode cooling. Displacement thermal noise spectra
observed close to the (1,7) mechanical resonance frequency, for
different frequency offsets $\Delta\omega$ between TEM$_{00}$ and
TEM$_{04}$ modes of the cavity. Due to the three-mode cooling, the
thermal noise spectrum is both widened and reduced. Curves (a) to (d) are measured for different detunings, (a) being the furthest from resonance, and (d) the closest (see Fig. \ref{fig:results}).}
\label{fig:cooling}

\end{figure}

\begin{figure}[t]
\begin{center}
\includegraphics[width=8.5cm]{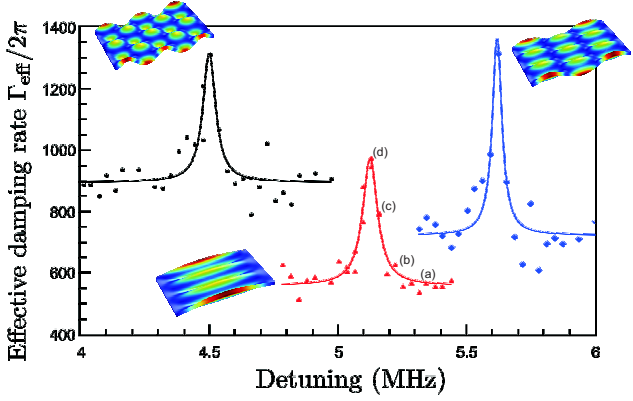}
\caption{Experimental results: demonstration of the resonant
character of the three-mode process. Mechanical dampings of the (1,7)
(red), (5,5) (black) and (3,7) (blue) mechanical modes. The damping
increases substantially close the resonance condition. Dots are
experimental values of the effective damping, while the lines are fits to
equation (\ref{eq:gainantiStokes}). The letters (a) to (d) refer to the spectra on Fig. \ref{fig:cooling}.
Inserts display the
simulated mode shape profiles of the mechanical modes.}
\label{fig:results}
\end{center}
\end{figure}

Experimental results for the anti-Stokes process are
presented in Fig. \ref{fig:cooling}, which displays the observed
noise spectra close to the (1,7) mechanical resonance frequency for
different values of $\Delta\omega$ (curves (a) to (d)). The corresponding detunings are shown in Fig. \ref{fig:results}.
 Curve (a) is taken for a detuning such that $|\Delta\omega-\Omega_{\rm m}|\gg
\Omega_{\rm cav}$, which results in the absence of any three-mode effect. Close to the anti-Stokes process
resonance ($\Delta\omega=+\Omega_{\rm m}$), the thermal noise
spectrum is both widened and reduced as expected. Similar effects were already observed in two-mode cooling experiments \cite{NatureLKB}. Curve (d) is taken close to the  resonance $\Delta\omega=\Omega_{\rm m}$
 and displays a parametric gain $\mathcal{R}\simeq -0.72$.

\begin{figure}[b!]
\begin{center}
\includegraphics[width=8.5cm]{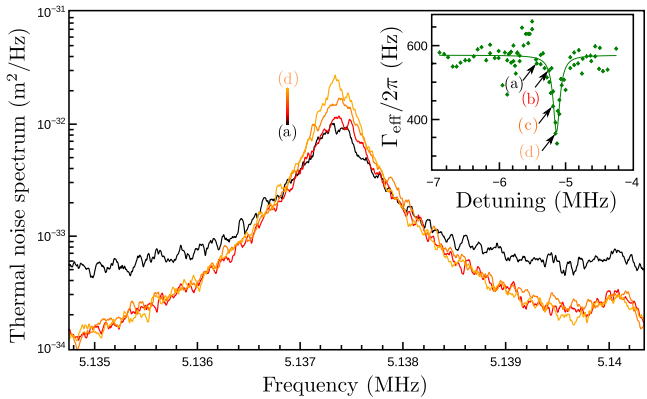}
\caption{Experimental demonstration of the Stokes mechanism. Thermal noise spectra observed close to the Stokes process resonance. The curves are for different detunings e to h, see insert.
 Insert: Evolution of the effective mechanical damping of the (1,7)
 mechanical mode close to the Stokes resonance condition, together with a fit with Eq. (\ref{eq:gainStokes}).}
\label{fig:resultsAmplification}
\end{center}
\end{figure}

Similar measurements have been carried out for the neighboring
mechanical modes (3,7) and (5,5) of the moving mirror. These results are also
presented in Fig. \ref{fig:results}. Here we have used the effective damping rate $\gamma_{\rm eff}$ of the mechanical resonator as a measure of the three-mode coupling strength. For all three modes, we find that the frequency detuning $\Delta\omega$ corresponding to the maximal damping effect  matches the mechanical resonance frequency $\Omega_{\rm m}$, deduced from the thermal noise spectrum, and the resulting damping/cooling of the mechanical mode. One can see that the mechanical damping goes back to its intrinsic value as soon as three-mode effects are negligible. Note that different clamping losses for each mode result in different intrinsic damping values. Despite the large dispersion due to the low stability of the cavity, experimental points for all three mechanical modes are well fitted by Eq. (\ref{eq:gainantiStokes}), with a common optical bandwidth value close to 40 kHz, of the same order of magnitude as the cavity optical bandwidth $\Omega_{\rm cav}$. Typical measured values for the absolute parametric gain are close to 0.5.

Three-mode amplification has also been demonstrated
 close to the Stokes process resonance $(\Delta\omega=-\Omega_{\rm m}$), as shown in Fig. \ref{fig:resultsAmplification}. Sweeping the cavity length over 40 $\mu$m to change the detuning by $2\Omega_{\rm m}/2\pi\simeq 10\,{\rm MHz}$ also causes change of the alignment of the cavity, which reduces the overlap $\Lambda$ and the observed parametric gain. Once again, results are in good agreement with Eq. (\ref{eq:gainStokes}), and the measured maximal value for the gain in this case is $\mathcal{R}=0.41$ for mode (1,7), for the same optical power as in the cooling experiment.
Even with an incident power up
to 5 mW, the three-mode parametric instability regime (corresponding to
$\mathcal R\geq1$) has eluded observation.

\section{Silicon Nitride Membrane in the Middle Cavity: Observation of the Parametric Instability\label{secExpUWA}}

To observe higher parametric gains, it is convenient to use a lower mass resonator. We have accordingly chosen a silicon nitride membrane as mechanical resonator. It has already been shown that such membranes have low optical absorption and can  be embedded in high-finesse optical cavities, creating a coupled cavity configuration that can mimick a conventional setup. They also have high quality factor mechanical modes in the MHz range \cite{Zwickl2008} .

\subsection{Experimental setup: design and characterization}

\begin{figure}[b]
\centering
\includegraphics[width=0.45\textwidth]{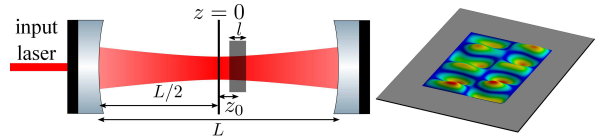}
\caption {'Membrane-in-the-middle' configuration for the three-mode interaction experiment. Left: Cavity configuration. Right: The (2,6) mechanical mode shape of the silicon nitride membrane supported by a silicon frame.}
\label{fig:cavity}
\end{figure}

Here we consider the system shown on Fig. \ref{fig:cavity}.
The membrane  is a commercial Norcada stoichiometric silicon nitride membrane with tensile stress $T=800\,\rm{Mpa}$ and density $\rho=2.7 \,\rm{g/cm^3}$ \cite{Craighead2006}. It has a square geometry with a side length  $D=1\,\rm{mm}$ and a thickness $l= 50\,\rm{nm}$. The effective mass of the membrane is $40\, \rm ng$.
The membrane is embedded close to the waist of a near confocal cavity.

\begin{figure}[t]
\centering
\includegraphics[width=0.5\textwidth]{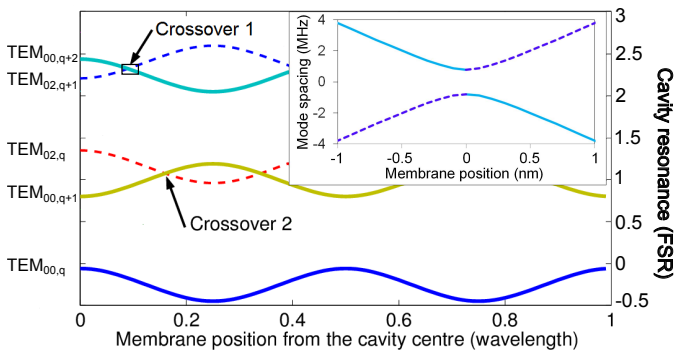}
\caption {Optical resonance frequencies of the compound cavity as a function of the membrane position $z_0$. For simplicity, only the relevant transverse modes TEM$_{00}$ and TEM$_{02}$ are shown. Insert: Close-up of the avoided crossing near Crossover 1 (with relative membrane position). Note the different horizontal and vertical scales.}
\label{fig:mode}
\end{figure}

The resonance frequency of such a cavity  depends on the reflectivity $r$ and relative position of the membrane $z_0$ \cite{membrane2008}:
\be
 \omega_q =\omega^0_q - (c/L)\cos^{-1}[|r|\cos(4\pi z_0/ \lambda)]\,,
\ee
where $\omega^0_q$ is the resonance frequency for the corresponding linear cavity, and  $r$ is the membrane amplitude reflectivity, which depends on the membrane thickness $l$ and index of refraction $n$ \cite{Brooker} as follows:
\be
\label{eq:memrefl}
r = \frac{(n^{2} -1)\sin 2\pi n l/\lambda}{2 i n \cos 2\pi n l/\lambda + (n^{2} +1)\sin 2\pi n l/\lambda}.
\ee
In this system, one also has to correct the overlap factors $\Lambda_{ij}$ by an additional dimensionless longitudinal overlap factor $\Lambda_{\rm l}$, which depends on the compound cavity workingpoint and is, in the limit $\abs {\omega_{\rm s}-\omega_{\rm p}} \ll \omega_{\rm p}$, given by:
\be
\label{eq:radpress-coeff-fin-diag}
\Lambda_{\rm l}=\sin(4\pi z_0/\lambda)\sqrt{\frac{ r}{1-\abs{r}^2\cos^2(4\pi z_0/\lambda)}}.
\ee

To demonstrate the parametric instability, the membrane position is tuned so that the two optical modes have a frequency difference $\Delta\omega$ equal to the mechanical resonance frequency $+\Omega_{\rm m}$.  Fig. \ref{fig:mode} shows the frequency structure of the coupled cavity as a function of the membrane position.  One can see that when the membrane is near Crossover 1 or 2, the frequency difference between the ${\rm TEM}_{00}$ mode and the target ${\rm TEM}_{02}$ mode can be small enough to match the membrane mechanical frequency.  However, Crossover 1 ($z_0 = 0.09\,\lambda$) is closer to a node of the electric field and should have lower optical loss due to membrane absorption \cite{Vitali2011} than Crossover 2 ($z_0 = 0.16\,\lambda$).

In principle, tuning the membrane position allows the mode spacing between the  ${\rm TEM}_{00}$ and  ${\rm TEM}_{20}$ modes to be tuned to arbitrarily small value to match the membrane mechanical resonance frequency. However in practice, the crossovers  are avoided  due to coupling between the modes \cite{OptLett, HarrisTuning}.  Thus there is a minimum frequency spacing at the nominal crossing point as shown in the insert in Fig. \ref{fig:mode}. One of the greatest challenges in this experiment has been the adjustment of the membrane position and its careful alignment to allow the minimum frequency spacing to be smaller than the chosen mechanical mode frequency.

\begin{figure}[b!]
\centering
\includegraphics[width=0.5\textwidth]{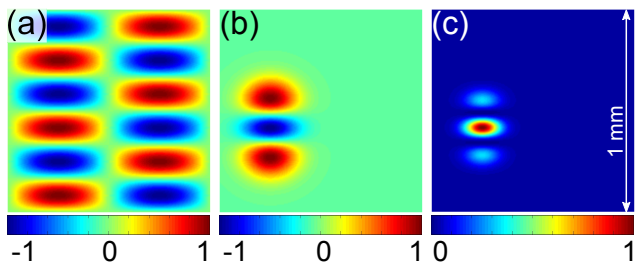}
\caption {Illustration of the overlap between the membrane $(2,6)$ mechanical  mode and $\rm TEM_{00}$ and  $\rm TEM_{20}$ optical modes. (a) Mode shape of the membrane (2,6) mode; (b) Mode shape of the ${\rm TEM}_{20}$ cavity mode. (c) Product of the mode shapes, with the optical mode correctly located on the membrane for optimised overlapping.}
\label{fig:modeshape}
\end{figure}

\begin{figure}[b]
\centering
\includegraphics[width=0.5\textwidth]{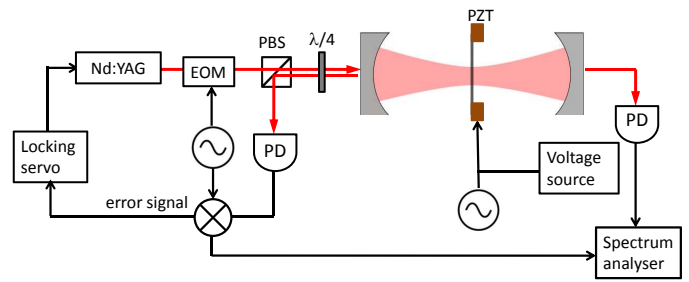}
\caption {Experimental setup  of the 'membrane-in-the-middle' configuration. A SiN membrane  is placed close to the cavity waist. An electro-optic phase modulator (EOM) and a polarized beam splitter (PBS) are used to lock the laser frequency to the cavity resonance using the PDH. The membrane is mounted on a PZT to tune its horizontal position. The transmitted light from the cavity is monitored by an offset photodiode (PD) in order to detect the beatnote between the fundamental mode and the higher-order mode.}
\label{fig:setup}
\end{figure}

Fig. \ref{fig:modeshape} shows the mode shapes of the (2,6) mechanical mode, the optical cavity  ${\rm TEM}_{20}$ mode shape profile, and the product of all three profiles - the (2,6) mode, the ${\rm TEM}_{00}$ mode and the ${\rm TEM}_{20}$ mode. It is clear that there is a good overlap between these modes if the cavity modes are correctly positioned on the membrane at a specific location. The optimized overlap factor is $\sim0.11$.

The experimental setup is shown in Fig. \ref{fig:setup}. The optical cavity is mounted on an invar bar in a vibration isolated vacuum tank. Motorized optical mounts and PZT are used for cavity alignment. The Pound-Drever-Hall (PDH) technique \cite{PDH} is used to lock the input laser frequency to the cavity ${\rm TEM}_{00}$ resonance. We have developed careful alignment and tuning procedures to tune the cavity. The cavity finesse is first measured  $\sim15,000$ without a membrane present.

The optical mode frequency spacing is set by tuning the membrane position and orientation.  The maximum finesse observed with the membrane inserted at Crossover 1 position is $\sim13,000$, corresponding to a cavity decay time of $1.38\, \mu{\rm  s}$.

\begin{figure}[b]
\centering
\includegraphics[width=0.5\textwidth]{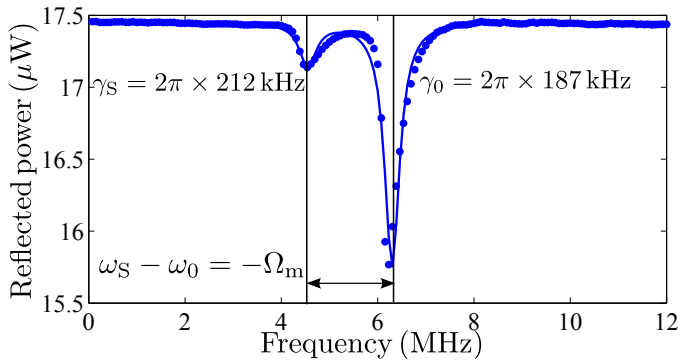}
\caption {Cavity spectrum when the mode spacing is tuned close to the mechanical mode frequency. A fit with a double Lorentzian allows the frequency difference $\Delta\omega$ and the cavity losses for both optical modes to be determined.}
\label{fig:gap}
\end{figure}

We have measured the dependence of ${\rm TEM}_{00}$ and ${\rm TEM}_{20}$ mode frequency spacing with the membrane angle. As expected, the frequency spacing is minimized when the membrane is normal to the cavity axis, and varies with a typical rate $\sim4\,{\rm MHz}/{\rm  mrad}$. We have first tuned the membrane position along the optical axis to the desired Crossover 1 position. The optical mode spacing at the avoided crossing is then tuned by membrane angular adjustment. Fig. \ref{fig:gap} shows the tuned cavity spectrum. We have fitted the spectrum to two Lorentzians to determine the corresponding mode linewidths. As expected, as the optical mode spacing decreases, the cavity mode linewidth increases due to the coupling between the two modes \cite {OptLett}.   The cavity is stable enough to make repeatable measurements, but  retuning is required every day to compensate for very slow drifts.

The measured  resonance frequency for the $(2,6)$ mode is $\Omega_{\rm m}/2\pi\sim 1.718\,{\rm MHz}$, consistent with the expected frequency $\Omega_{i,j}/2\pi = \sqrt{T/4 \rho D^2}\sqrt{i^2+j^2}$. For a stressed membrane, the normalized membrane $(i,j)$ mode shape is  given as follows $(i,j=1,2,...)$ \cite{Vitali2011}:
\be
u_{i,j}(x,y) = \frac{2}{D}\sin\left(\frac{i\pi x}{D}+\frac{i\pi}{2}\right)\sin\left(\frac{j\pi y}{D}+\frac{j\pi}{2}\right).
\ee
The mode shape of the $(2,6)$ mode  is shown in Fig. \ref{fig:modeshape}.
At a pressure of $10^{-4}\,{\rm mbar}$, the membrane has a mechanical decay time of 185 ms, corresponding to  a Q factor of $\simeq 10^6$.

\subsection{Experimental results with the membrane-in-the-middle cavity}

\begin{figure}[b]
\centering
\includegraphics[width=0.5\textwidth]{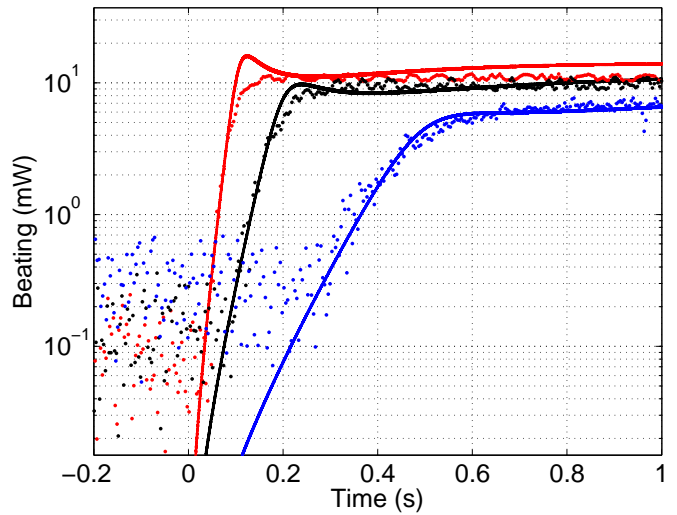}
\caption {Experimental results with relevant simulations. The dots are experimental data of the beatnote signal between the cavity ${\rm TEM}_{00}$ mode and ${\rm TEM}_{20}$ mode at mechanical frequency, as a function of time for different input powers. The curves display exponential ring-up before reaching saturation. The solid curves are the simulation results for the corresponding input power.  The input power is $26\,\mu{\rm  W}$ (red), $15\,\mu{\rm  W}$ (black) and $7\,\mu {\rm  W}$ (blue).}
\label{fig:timefit}
\end{figure}

Once the cavity is correctly tuned, the laser is locked to the TEM$_{00}$ mode and exponential ring-up of the mechanical $(2,6)$ mode occurs as soon as the input power exceeds the threshold of $R=1$. Consistent with the results in Sec. \ref{secModel}, the exponential ring-up reaches saturation and does not ruin the cavity locking.  The Stokes mode amplitude reaches saturation when it approaches the cavity linear dynamic range $\lambda/\mathcal{F}\sim 10^{-10} {\rm m}$.  This occurs in a time of between  0.1 and 0.5 seconds, as shown in Fig. \ref{fig:timefit}. The figure shows that the saturation amplitude depends on the input power. Due to the very low mass membrane and the high-finesse cavity used in this experiment, the threshold for parametric instability is expected to be only $\sim 3 \,\mu \rm W$. Hence the experiments have to be conducted at very low optical power, which means that the beatnote signal between ${\rm TEM}_{00}$ and ${\rm TEM}_{20}$  is below the photodetector noise floor until the mechanical amplitude has built up.

\begin{figure}[t]
\centering
\includegraphics[width=0.5\textwidth]{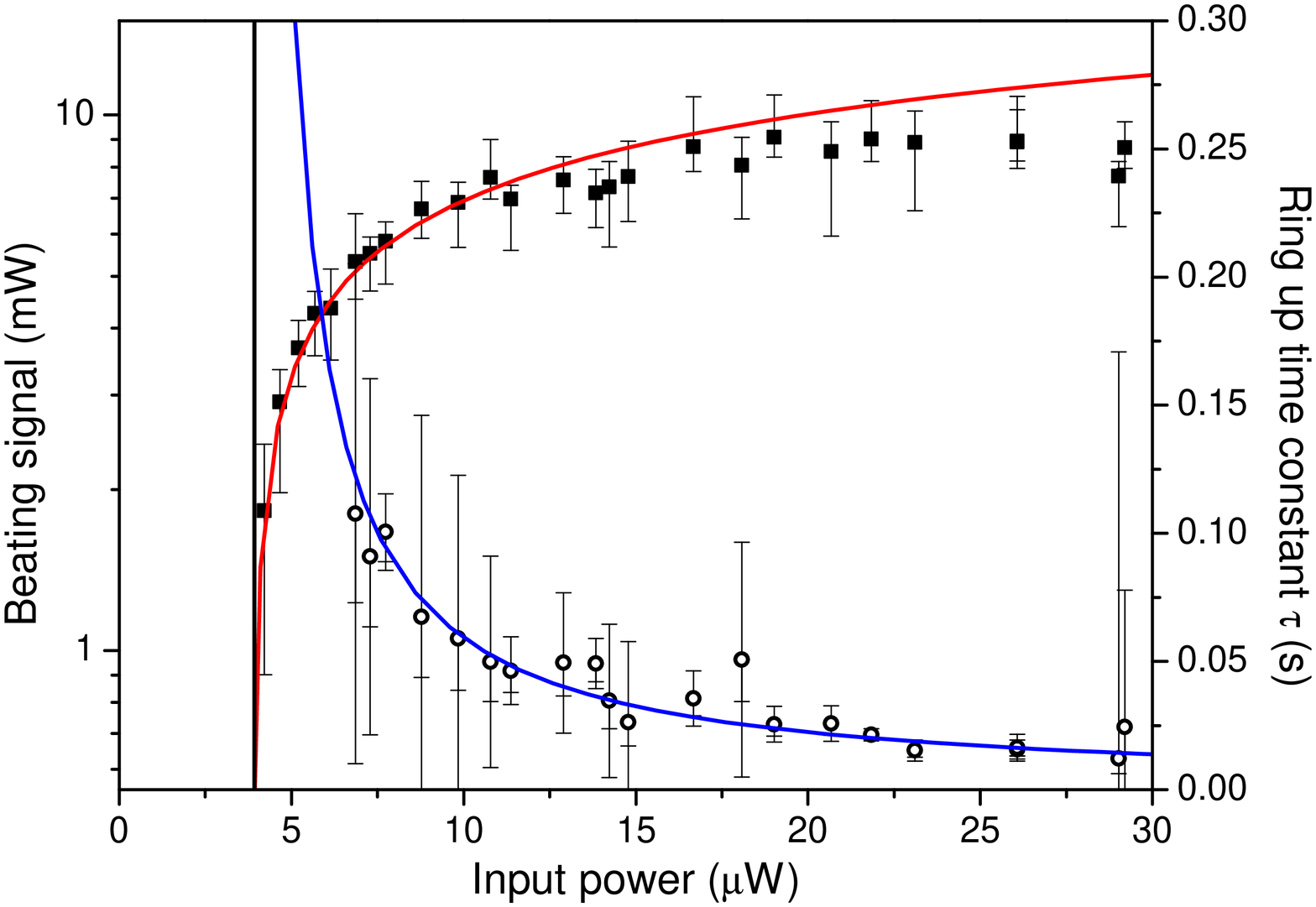}
\caption {Measured steady-state beatnote signal amplitudes (solid squares) and  ring-up times (hollow circles) as a function of the input power. The blue full line fit for the ring-up time is from Eq.  \eqref{EqGamma}, and yields a value of $G_{\rm ps}=2\pi\times0.10$ Hz. The red full line fit for the beatnote amplitude is from Eqs. \eqref{eq:steady_sol_a0} and \eqref{eq:steady_sol_as}.  The vertical line indicates the parametric instability threshold of $3.92\,\mu$W.}
\label{fig:taufit}
\end{figure}

Experimental results are presented on Fig. \ref{fig:timefit} for three different optical input powers: 7, 15 and 26 $\mu$W, while the solid curves are simulations using the value of $G_{\rm ps}$ inferred from the fit performed in Fig. \ref{fig:taufit} and the same input powers.
The measured ring-up times and beatnote steady-state amplitudes between the ${\rm TEM}_{00}$ and ${\rm TEM}_{20}$ modes are shown in Fig. \ref{fig:taufit} as a function of input power. The curves display an excellent agreement between the experimental results and the large-amplitude model, except for some discrepancy at large input power. This could be due to thermal effects caused by the increased transverse mode intensity, which could lead to a decrease of the parametric gain. The fit for the ring-up time constant yields a value of $G_{\rm ps}=2\pi\times0.10\,\rm{Hz}$ and a corresponding instability threshold power $P_{\rm in}^{\rm th} = 3.92\, \mu\mathrm{W}$ using Eq. \eqref{eq:threshold}. This is in excellent agreement with the theoretical model as it corresponds to an effective coupling and an effective overlap very close to its optimum value $\Lambda = 0.95\,\Lambda_{\rm opt}$, which can be explained by the imperfect  membrane alignment.
The fit for the steady-state beatnote amplitude yields a similar value.

 \section*{Conclusion}

 We have designed and operated two different tunable free-space optomechanical systems which demonstrate three-mode parametric effects, either on the Stokes or the anti-Stokes sideband. With a bridge resonator, we have demonstrated radiation-pressure cooling  and optomechanical amplification of the resonator Brownian motion, and  the resonant character of the three-mode optomechanical effects. Using a membrane resonator, we have observed three-mode parametric instability in a free-space optical cavity. Unlike other recent related experiments, both our systems are fully tunable and can be used to fully test the relevant theory.

We have shown that the theory of  three-mode optomechanical interactions correctly predicts the onset of parametric instability. The time-dependence of parametric instability is in quantitative agreement with the original theory of Braginsky. We find that mechanical mode amplitudes saturate in accordance with our large-amplitude model for parametric instability. The saturation acts to control the cavity pump mode power to a constant value independent of the input laser power.
The very low threshold power for parametric instability confirms the intrinsic efficiency of three-mode optomechanical systems, supporting predictions  that three-mode systems could be very effective tools for ground state cooling \cite{zhaoPRL2009}.

In the experiment performed here, the onset of three-mode instability has not led to loss of cavity lock as the power lose from the main cavity mode is sufficient to stabilize the system. If the same behavior happens in high-power laser interferometers for gravitational-wave detection, it should be much easier to implement instability control techniques based on feedback or slow thermal tuning.

{\bf Acknowledgements}

We wish to thank the Gingin Advisory Committee of the LIGO Scientific Collaboration, the LIGO Scientific Collaboration Optics Working Group and our collaborators Stefan Gossler, Gregg Harry, Stan Whitcomb and Samuel Del\'eglise for encouragement and useful advice. This research was supported by the Australian Research Council, by the ¡°Agence Nationale de la Recherche¡±
program ¡°ANR-2011-B504-028-01 MiNOToRe,¡± and by the European FP7 Specific Targeted Research Projects Minos and QNEMS.


\begin{thebibliography}{1}
\bibitem{Braginsky2001}  V. B. Braginsky, S. E. Strigin, and S. P. Vyatchanin, Phys. Lett. A {\bf 287}, 331 (2001).
\bibitem{Braginsky2002}  V. B. Braginsky, S. E. Strigin, and S. P. Vyatchanin, Phys. Lett. A {\bf 305}, 111 (2002).
\bibitem{zhao2005} C. Zhao, L. Ju, J. Degallaix, S. Gras, and D. G. Blair, Phys. Rev. Lett. {\bf 94},121102 (2005).
\bibitem{KellsMod} W. Kells, LIGO document LIGO-G070145-x0 (2007).
\bibitem{Evans} M. Evans, L. Barsotti, and P. Fritschel, Phys. Lett. A {\bf 374}, 665 (2010).
\bibitem{zhaoPRA2008} C. Zhao {\it et al.}, Phys. Rev. A {\bf 78}, 023807 (2008).
\bibitem{zhaoPRA2011} C. Zhao \emph{et al.}, Phys. Rev. A {\bf 84}, 063836 (2011).
\bibitem{Grudinin} I.~S. Grudinin, H. Lee, O. Painter, and K.~J. Vahala, Phys. Rev. Lett. {\bf 104}, 083901 (2010).
\bibitem{Bahl}  G. Bahl, J. Zehnpfennig, M. Tomes, and T. Carmon, Nat. Commun. {\bf 2}, 1038 (2011).
\bibitem{Brillouin1965} Y.~R. Shen and N. Bloembergen, Phys. Rev. {\bf 137}, A1787 (1965).
\bibitem{Kippenberg2010} J.~M. Dobrindt and T.~J. Kippenberg, Phys. Rev. Lett. {\bf 104}, 033901 (2010).
\bibitem{MIT1} T. Corbitt, D. Ottaway, E. Innerhofer, J. Pelc, and N. Mavalvala, Phys. Rev. A {\bf 74}, 021802 (2006).
\bibitem{Kippenberg1} T. J. Kippenberg, H. Rokhsari, T. Carmon, A. Scherer, and K. J. Vahala, Phys. Rev. Lett. {\bf 95}, 033901 (2005).
\bibitem{Kippenberg2} H. Rokhsari, T. J. Kippenberg, T. Carmon, and K. J. Vahala, Opt. Express {\bf 13}, 5293 (2005).
\bibitem{Kippenberg3} P. Del'Haye, A. Schliesser, O. Arcizet, T. Wilken, R. Holzwarth, and T. Kippenberg, Nature (London), {\bf 450}, 1214 (2007).
\bibitem{Ma} R. Ma, A. Schliesser, P. Del'Haye, A. Dabirian, G. Anetsberger, and T.~J. Kippenberg, Opt. Lett. {\bf 32}, 2200 (2007).
\bibitem{Carmom2007} T. Carmon and K. J. Vahala, Phys. Rev. Lett. {\bf 98}, 123901 (2007).
\bibitem{Carmon} T. Carmon, H. Rokhsari, L. Yang, T. J. Kippenberg, and K. J. Vahala, Phys. Rev. Lett. {\bf 94}, 223902 (2005).
\bibitem{CarmonBahl} G. Bahl, M. Tomes, F. Marquardt, and T. Carmon, Nat. Phys. {\bf 8}, 203 (2012).
\bibitem{Grudinin2009} I. S. Grudinin, A. B. Matsko, and L. Maleki, Phys. Rev. Lett. {\bf 102}, 043902 (2009).
\bibitem{Tomes2009} M. Tomes and T. Carmon, Phys. Rev. Lett. {\bf 102}, 113601 (2009).
\bibitem{SAW2009} A. B. Matsko, A. A. Savchenkov, V. S. Ilchenko, D. Seidel, and L. Maleki, Phys. Rev. Lett. {\bf 103}, 257403 (2009).
\bibitem{SAW2011} A. A. Savchenkov, A. B. Matsko, V. S. Ilchenko, D. Seidel, and L. Maleki, Opt. Lett. {\bf 36}, 3338 (2011).
\bibitem{Kippenberg4} G. Anetsberger, E.M. Weig, J.P. Kotthaus and T.J. Kippenberg, C. R. Physique {\bf 12},  800 (2011).
\bibitem{Tobar1} M. E. Tobar and D. G. Blair, J. Phys. D: Appl. Phys. {\bf 26}, 2276 (1993).
\bibitem{aLIGO} https://www.advancedligo.mit.edu/
\bibitem{aVirgo} J. Degallaix  \emph{et al.}, Astronomical Society of the Pacific  Conference Series, 9th LISA Symposium {\bf 467}, 151 (2013).
\bibitem{NatureLKB} O. Arcizet, P.-F. Cohadon, T. Briant, M. Pinard, and A. Heidmann, Nature (London) {\bf 444}, 71 (2006).
\bibitem{Zwickl2008} B. M. Zwickl \emph{et al.}, Appl. Phys. Lett. {\bf 92}, 103125 (2008).
\bibitem{Kippenberg} A. Schliesser, O. Rivi\`ere, G. Anetsberger, O. Arcizet, and T.J. Kippenberg, Nat. Phys. {\bf 4}, 415 (2008).
\bibitem{TheorySelfCool} I. Wilson-Rae, N. Nooshi, W. Zwerger and T.J. Kippenberg, Phys. Rev. Lett. \textbf{99}, 093901 (2007).
\bibitem{TheorySelfCool2} F. Marquardt, J.P. Chen, A.A. Clerk and S.M.Girvin, Phys. Rev. Lett. \textbf{99}, 093902 (2007).
\bibitem{QGS_Teufel} J.D. Teufel \emph{et al.}, Nature (London) \textbf{475}, 359 (2011).
\bibitem{QGS_Painter} J. Chan \emph{et al.}, Nature (London) \textbf{478}, 89 (2011).
\bibitem{QGS_TJK} E. Verhagen, S. Del\'eglise, S. Weis, A. Schliesser, and T.J. Kippenberg, Nature (London) \textbf{482}, 63 (2012).
\bibitem{Juli3mode} L. Ju, S. Gras, C. Zhao, J. Degallaix, and D.G. Blair, Phys. Lett. A {\bf 354}, 360 (2006).
\bibitem{Kells} W. Kells and E. D'Ambrosio, Phys. Lett. A {\bf 299}, 326 (2002).
\bibitem{Strigin} S.E. Strigin and S.P. Vyatchanin, Phys. Lett. A {\bf 365}, 10 (2007).
\bibitem{Vitali2011} C. Biancofiore \emph{et al.}, Phys. Rev. A {\bf 84}, 033814 (2011).
\bibitem{Press1992} W. H. Press, B. P. Flannery, S. A. Teukolsky, and W. T. Vetterling, \emph{Numerical Recipes in FORTRAN: The Art of Scientific Computing} (Cambridge University Press, Cambridge, 1992), pp. 704-716.
\bibitem{PRL_Arcizet} O. Arcizet \emph{et al.}, Phys. Rev. Lett.  \textbf{97}, 133601 (2006).
\bibitem{membrane2008} A. M. Jayich \emph{ et al.}, New J. Phys. {\bf 10}, 095008 (2008).
\bibitem{Brooker}G. Brooker, \emph{Modern Classical Optics} (Oxford University Press, Oxford, 2003).
\bibitem{OptLett} T. Klaassen, J. de Jong, M. van Exter, and J. P. Woerdman, Opt. Lett. {\bf 30}, 1959 (2005).
\bibitem{HarrisTuning} J. C. Sankey, C. Yang, B. M. Zwickl, A. M. Jayich, and J. G. E. Harris, Nature Phys. {\bf 6}, 707 (2010).
\bibitem{PDH}  R. W. P. Drever \emph{et al.}, Appl. Phys. B {\bf 31}, 97 (1983). E. D. Black, Am. J. Phys. {\bf 69}, 79 (2001).
\bibitem{Craighead2006} S. S. Verbridge, J. M. Parpia, R. B. Reichenbach, L. M. Bellan, and H. G. Craighead, J. Appl. Phys. \textbf{99}, 124304 (2006).
\bibitem{zhaoPRL2009} C. Zhao, L. Ju, H. Miao, S. Gras, Y. Fan, and D. G. Blair, Phys. Rev. Lett {\bf 102}, 243902  (2009).




\end{thebibliography}
\end{document}